# Observation of the Cabibbo Suppressed Charmed Baryon Decay $\Lambda_c^+ \to p\phi$

**Abstract**

We report the observation of the Cabibbo-suppressed decays $\Lambda_c^+ \to pK^+K^-$ and $\Lambda_c^+ \to p\phi$ using data collected with the CLEO II detector at CESR. The latter mode, observed for the first time with significant statistics, is of interest as a test of color-suppression in charm decays. We have determined the branching ratios for these modes relative to $\Lambda_c^+ \to pK^-\pi^+$ and compared our results with theory.

The strength of color-suppression in internal W-emission charmed meson decays has long been in question. For example, $B(D_s^+ \to \bar{K}^{*0}K^+)/B(D_s^+ \to \phi\pi^+) \simeq 1$ [1,2], while the expectation from color-matching requirements is that this ratio should be about 1/18. Reasonable overall agreement with the experimental data in the charm sector has been obtained using factorization and taking the large $N_c$ limit in a $1/N_c$ expansion approach, where $N_c$ is the number of quark colors [3,4]. The Cabibbo-suppressed charmed baryon decay $\Lambda_c^+ \to p\phi$, shown in Figure 1, is also naively expected to be color-suppressed. However, using factorization and taking the limit $N_c \to \infty$ leads to a prediction of no color-suppression [5]. The $\Lambda_c^+ \to p\phi$ decay receives contributions only from factorizable diagrams, and therefore a reliable calculation should be obtained using factorization. Observation of the decay mode $\Lambda_c^+ \to p\phi$ was first reported by the ACCMOR collaboration with $2.8 \pm 1.9$ events [9]. Last year the E687 collaboration published results on the first observation of the Cabibbo-suppressed charmed baryon decay $\Lambda_c^+ \to pK^+K^-$, along with an upper limit on the resonant substructure $\Lambda_c^+ \to p\phi$ [10]. Herein we present new CLEO results on the observation of $\Lambda_c^+ \to pK^+K^-$ and $\Lambda_c^+ \to p\phi$ decays and discuss the implications of the results.

We use a data sample recorded with the CLEO II detector operating at the Cornell Electron Storage Ring (CESR). The sample consists of $e^+e^-$ annihilations taken at and slightly below the $\Upsilon(4S)$ resonance, for a total integrated luminosity of 3.46 fb$^{-1}$. The main detector components which are important for this analysis are the tracking system and the barrel Time-of-Flight (TOF) particle identification system. Additional particle ID is provided by specific ionization ($dE/dx$) information from the tracking system's main drift chamber. A more detailed description of the CLEO II detector has been provided elsewhere [11].

To search for the $\Lambda_c^+$ signals, we study $pK^+K^-$ track combinations found by the tracking system. The $p$ and $K^\pm$ candidates are identified by combining information from the TOF and dE/dx systems to form a combined $\chi^2$ probability $\mathcal{P}_i$ for each mass hypothesis $i = \pi, K, p$. Using these probabilities $\mathcal{P}_i$, a normalized probability ratio $L_i$ is evaluated for each track according to the formula: $L_i \equiv \mathcal{P}_i/(\mathcal{P}_\pi + \mathcal{P}_K + \mathcal{P}_p)$. Well-identified protons form a sharp peak near $L_p = 1$, while tracks identified as *not* being protons form a peak near $L_p = 0$. The remainder of the candidates fall in the region between 0 and 1. For the proton involved in each decay mode under study, we require $L_p > 0.9$, which constitutes a strong cut. For the kaons, we apply a loose cut of $L_K > 0.1$. In addition, all protons and kaons must pass a minimum requirement of $\mathcal{P}_p > 0.001$ and $\mathcal{P}_K > 0.001$, respectively. In order to reduce the large combinatoric background, the candidate $\Lambda_c^+$ scaled momentum $x_p = P_{\Lambda c}/\sqrt{E_{beam}^2 - m_{\Lambda c}^2}$ is limited to $x_p > 0.5$.

The $pK^+K^-$ invariant mass is shown in Figure 2. The broad enhancement in the mass region above 2.37 GeV/c$^2$ is a reflection from the decay mode $\Lambda_c^+ \to pK^-\pi^+$, the pion has been misidentified as a kaon. The spectrum is fitted to a Gaussian for the signal with width fixed to $\sigma = 4.9$ MeV/c$^2$ determined from Monte Carlo simulation [12], and a 2$^{nd}$ order Chebychev polynomial for the smooth background. This fit yields $214 \pm 50$ events for the inclusive $\Lambda_c^+ \to pK^+K^-$ signal with a mean mass of $2285.5 \pm 1.2$ MeV/c$^2$ [13].

To find a $\Lambda_c^+ \to p\phi$ signal, we reconstruct $\phi$ candidates through their decays $\phi \to K^+K^-$. Because the width of the $\phi$ is comparable to the detector mass resolution, the $\phi$ signal shape is best described by a convolution of a Gaussian and a Breit-Wigner of width $\Gamma = 4.43$ MeV/c$^2$ [1]. The background is parameterized by a function of the form $b(m) = N(m-m_0)^\alpha e^{\beta(m-m_0)}$. The measured Gaussian resolution from the fit is $\sigma = 1.6 \pm 0.2$ MeV/c$^2$. In order to perform background subtractions, $1.0121 < m_{KK} < 1.0273$ GeV/c$^2$ is designated as the $\phi$ "signal" region, while $0.990 < m_{KK} < 1.005$ GeV/c$^2$ and $1.035 < m_{KK} < 1.050$ GeV/c$^2$ are designated as the "sideband" regions. Integrating the background function over the sideband and signal regions gives a signal-to-sideband scale factor $R_\phi = 0.560 \pm 0.016$, which is used in the $\phi$ background subtraction below.

In order to obtain the $\Lambda_c^+ \to p\phi$ signal, the $pK^+K^-$ mass plot is made both for $m_{K^+K^-}$ in the $\phi$ signal region and the $\phi$ sideband regions. Figure 3 shows the results. The spectra are fitted to a Gaussian for the signal with width fixed to $\sigma = 4.9$ MeV/c$^2$ from Monte Carlo, and a 2nd order Chebychev polynomial for the smooth background. The fit to the $pK^+K^-$ mass spectrum corresponding to the $\phi$ signal region yields $54 \pm 12$ events with a confidence level of 97%. The mean mass for the signal is measured to be $2288.2 \pm 1.3$ MeV/c$^2$. In fitting the $pK^+K^-$ mass corresponding to the $\phi$ sideband region, the mean $\Lambda_c^+$ mass is fixed to that obtained from the $\phi$ signal region and the $\sigma$ is fixed to the Monte Carlo value as before. This gives $-16.4 \pm 9.6$ events for the $\phi$-sideband $\Lambda_c^+$ yield. Since the true contribution must be positive-definite, we set the central value to zero and use $0. \pm 9.6$ as the best estimate of the $\Lambda_c^+ \to pK^+K^-$ contribution. After scaling this by $R_\phi$ and subtracting, we find that the net $\Lambda_c^+ \to p\phi$ yield is $54 \pm 13$ events. This gives $-16.4 \pm 9.6$ events for the $\phi$-sideband $\Lambda_c^+$ yield. Since the true contribution must be positive-definite, we set the central value to zero and use $0. \pm 9.6$ as the best estimate of the $\Lambda_c^+ \to pK^+K^-$ contribution. After scaling this by $R_\phi$ and subtracting, we find that the net $\Lambda_c^+ \to p\phi$ yield is $54 \pm 13$ events. *[ NOTE TO CLEO READERS: Please See the CLEO Appendix at the end of this document.]*

As a check of the non-resonant contribution to the $\Lambda_c^+ \to p\phi$ signal, we fit the $K^+K^-$ mass spectra corresponding to the $\Lambda_c^+$ signal and sideband regions as determined from the inclusive $pK^+K^-$ mass spectrum. The $\phi$ yield obtained from the $\Lambda_c^+$ sideband regions, $2.246 < m_{pKK} < 2.266$ and $2.306 < m_{pKK} < 2.326$ GeV/c$^2$, is subtracted from that for the $\Lambda_c^+$ signal region, $2.276 < m_{pKK} < 2.296$ GeV/c$^2$. Figure 4 shows the fits to the $K^+K^-$ spectra from the $\Lambda_c^+$ signal and sideband regions, which yield $\phi$ signals of $92.2 \pm 17.0$ events and $36.5 \pm 13.5$ events, respectively. The $\Lambda_c^+$ sideband $K^+K^-$ mass spectrum in the Figure 4 has been scaled by the $\Lambda_c^+$ signal-to-sideband scale factor of $R_{\Lambda_c^+} = 0.502 \pm 0.013$, obtained by integrating the background function in Figure 2 over the $\Lambda_c^+$ signal and sideband regions. This gives $56 \pm 22$ events for the $\Lambda_c^+ \to p\phi$ signal, which is in agreement with the first method.

A check is also made for a possible reflection from $D_s^+ \to \phi\pi^+$, where the pion is misidentified as a proton. It is found that the reflection is a broad enhancement in the mass region above the signal. The effect of this background is minimized by the tight particle-ID requirement on the proton. Consequently, the overall fake rate is less than 1%, causing negligible effect on the $\Lambda_c^+ \to p\phi$ signal yield from the fit.

The decay $\Lambda_c^+ \to pK^-\pi^+$ is used as the normalization mode for the $\Lambda_c^+ \to p\phi$ relative branching ratio. In finding the $\Lambda_c^+ \to pK^-\pi^+$ yield, the same cuts are applied as in the $\Lambda_c^+ \to pK^+K^-$ analysis to minimize systematic errors, except that the particle-ID for the $\pi^+$ is loosened to a consistency requirement: $\mathcal{P}_\pi > 0.001$. The $\Lambda_c^+ \to pK^-\pi^+$ mass spectrum is shown in Figure 5. The parameterization of the fit is the same as the $\Lambda_c^+ \to p\phi$ mass fit in Figure 3, except that the width of the Gaussian is allowed to vary. The fit yields $5683 \pm 138$ observed signal events with a mean of $2286.8 \pm 0.2$ MeV/c$^2$ and a width of $6.4 \pm 0.2$ MeV/c$^2$. If the width of the Gaussian is fixed to the Monte Carlo prediction of 5.8 MeV/c$^2$, the yield changes by 4%. This dependence is included in the systematic error.

Monte Carlo simulation is used to determine all aspects of the detection efficiency except particle-ID. The particle-ID efficiency for protons is obtained using a sample of 33000 $\Lambda \to p\pi^-$ decays with a signal-to-background ratio of 50:1 [16]. For protons thus identified, the momentum spectrum after the particle-ID cuts ($L_p > 0.9$, $\mathcal{P}_p > 0.001$) is divided by the momentum spectrum before these cuts, bin by bin, yielding the particle-ID efficiencies versus momentum. To calculate the detection efficiency, the measured efficiency is folded in by randomly rejecting the corresponding fraction of Monte Carlo tracks in each momentum bin. The particle-ID ($L_K > 0.1$, $\mathcal{P}_K > 0.001$) efficiency for the kaons is derived in an analogous manner, except that the kaons are taken from $D^*$ decays through the cascade process $D^{*+} \to D^0 \pi^+$, $D^0 \to K^- \pi^+$. A sample of 11000 such $D^0 \to K^- \pi^+$ decays is obtained with an 8:1 signal-to-background ratio [16]. The particle-ID efficiency for protons is near 90% from 300 MeV/c$^2$ to 1.1 GeV/c$^2$ falling off to below 10% by 2.5 GeV/c$^2$. For kaons the particle-ID efficiency remains relatively flat at about 95%.

Using a Monte Carlo sample of $\Lambda_c^+ \to p\phi$ decays, where the $\Lambda_c^+$ fragmentation takes place according to the Lund JETSET Monte Carlo [17], the full detection efficiency is determined, with the particle-ID portion folded in as described above. For $\Lambda_c^+ \to p\phi$, the overall efficiency is 0.178±0.004 including the particle-ID efficiency which is 0.425±0.011. For $\Lambda_c^+ \to pK^+K^-$ (non-resonant) and $\Lambda_c^+ \to pK^-\pi^+$ the overall efficiencies are 0.216±0.005 and 0.224±0.005, respectively.

Since for all the decay modes the requirement $x_p > 0.5$ is applied, the relative branching ratio for each mode is found simply by dividing the corrected yields. Table I gives the details, listing only the statistical errors. The $\phi \to K^+K^-$ branching ratio is explicitly included in the calculation of the $\Lambda_c^+ \to p\phi$ branching ratio, and its uncertainty is included in the systematic errors.

The estimates for the main sources of systematic error include the $\Lambda_c^+ \to p\phi$ and $\Lambda_c^+ \to pK^+K^-$ signal shapes (7% and 11%, respectively) and background shapes (2% and 10%, respectively), particle-ID efficiency (6%), and the $\Lambda_c^+ \to pK^-\pi^+$ fit (4%). In addition, for the $\Lambda_c^+ \to p\phi$ mode, varying the $\phi$ signal and sideband regions gives a 5% variation in the yield. Finally, there is a 1.8% contribution to the $\Lambda_c^+ \to p\phi$ systematic error from the $\phi \to K^+K^-$ branching ratio uncertainty. Thus we estimate 12% systematic error in B($p\phi/pK\pi$), 17% in B($pKK/pK\pi$), and 18% in B($p\phi/pKK$). The final results appear in Table II, along with those from NA32 [9] and E687 [10]. Also shown in the table are theoretical predictions from Cheng and Tseng [5], Körner and Krämer [6], Żenczykowski [7], and Datta [8].

In summary, we have observed the Cabibbo-suppressed decays $\Lambda_c^+ \to p\phi$ and $\Lambda_c^+ \to pK^+K^-$. The results appear in Table II, which show that the phenomenological treatments of the $\Lambda_c^+ \to p\phi$ decay rate agree within a factor of two or three with our result. Since naive color suppression arguments would have required a $\Lambda_c^+ \to p\phi$ decay rate much smaller (by a factor of about 15), our result suggests, at least for the factorizable diagrams, that color-suppression is inoperative in charm baryon decays [5]. Furthermore, within the factorization approach using a $1/N_c$ expansion, our result supports the validity of taking the large $N_c$ limit in charm baryon decays.

R. Brun *et al.*, GEANT 3.14, CERN DD/EE/84-1.

[13] The quoted uncertainties in mass measurements refer to statistical error only.

[14] The remaining background is removed by sideband subtraction.

[15] T. Sjöstrand, Comp. Phys. Comm., **43** 367 (1987).

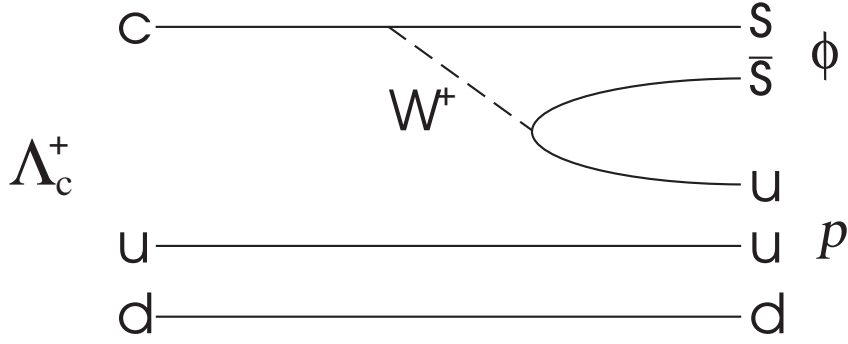

FIG. 1. The decay $\Lambda_c^+ \to p\phi$.

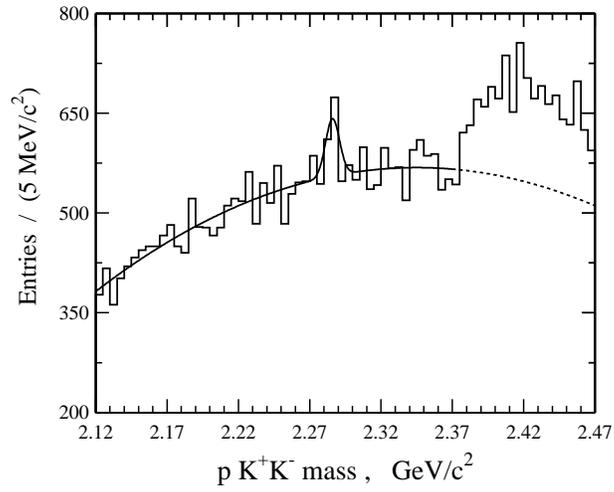

FIG. 2. Invariant mass of inclusive $pK^+K^-$ combinations passing all requirements. No $\phi$ cut is applied. The region above 2.37 GeV/c$^2$, where there is a large enhancement from $\Lambda_c^+ \to pK^-\pi^+$ decays, is not included in the fit.

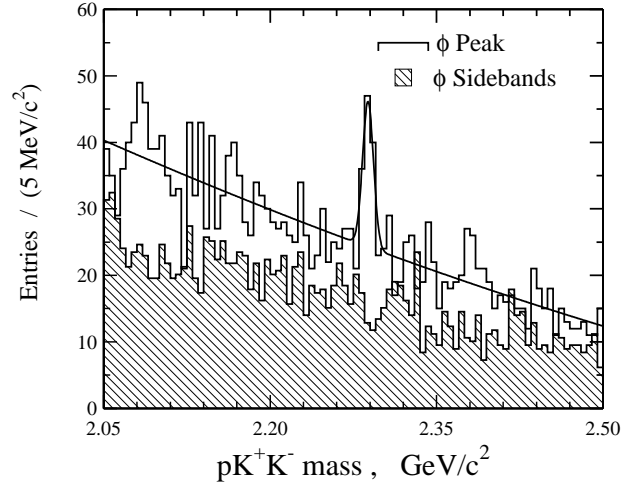

FIG. 3. Invariant mass of $pK^+K^-$ combinations corresponding to $K^+K^-$ mass in the $\phi$ "signal" (unshaded) and "sideband" (shaded) regions.

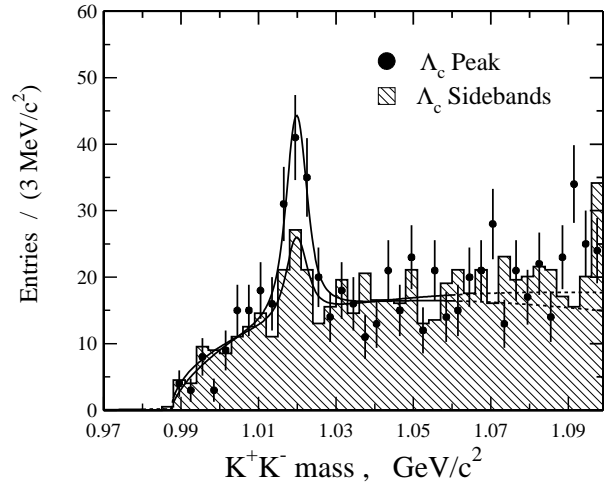

FIG. 4. Fit to $K^+K^-$ mass from combinations belonging to the $\Lambda_c^+$ signal (unshaded) and sideband (shaded) regions. The region above 1.06 GeV/c$^2$ is not included in the fit because of $K^{*0}$ feed-up when the $\pi$ is misidentified as a $K$.

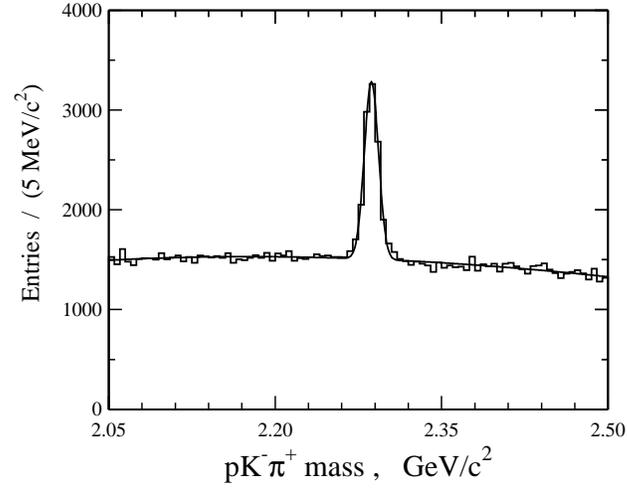

FIG. 5. Invariant mass of $pK^-\pi^+$ combinations found in the same data sample. The $\Lambda_c^+ \to pK^-\pi^+$ signal is used for normalization of the $\Lambda_c^+ \to p\phi$ branching ratio.

TABLE I. Calculation of the branching ratios for $\Lambda_c^+ \to p\phi$ and $\Lambda_c^+ \to pK^+K^-$ relative to $\Lambda_c^+ \to pK^-\pi^+$ and $\Lambda_c^+ \to pK^+K^-$. The errors are statistical only.

| $\Lambda_c^+$ Decay Mode: | $\Lambda_c^+ \to p\phi$ | $\Lambda_c^+ \to pK^+K^-$ | $\Lambda_c^+ \to pK^-\pi^+$ |
|---|---|---|---|
| Raw Yield | $54 \pm 12$ | $214 \pm 50$ | $5683 \pm 138$ |
| Efficiency | $0.178 \pm 0.004$ | $0.216 \pm 0.005$ | $0.224 \pm 0.005$ |
| $B(\phi \to K^+K^-)$ | $0.491 \pm 0.005$ | | |
| Corrected Yield | $618 \pm 138$ | $991 \pm 233$ | $25371 \pm 837$ |
| $B/B(\Lambda_c^+ \to pK^-\pi^+)$ | $0.024 \pm 0.006$ | $0.039 \pm 0.009$ | 1 |
| $B/B(\Lambda_c^+ \to pK^+K^-)$ | $0.62 \pm 0.20$ | 1 | |

TABLE II. Final results on $\Lambda_c^+ \to p\phi$ and $\Lambda_c^+ \to pK^+K^-$.

| Ratio of interest: | $\mathcal{B}(p\phi)/\mathcal{B}(pK\pi)$ | $\mathcal{B}(pKK)/\mathcal{B}(pK\pi)$ | $\mathcal{B}(p\phi)/\mathcal{B}(pKK)$ |
|---|---|---|---|
| This experiment | $0.024 \pm 0.006 \pm 0.003$ | $0.039 \pm 0.009 \pm 0.007$ | $0.62 \pm 0.20 \pm 0.12$ |
| NA32 | $0.04 \pm 0.03$ | | |
| E687 | | $0.096 \pm 0.029 \pm 0.010$ | $< 0.58@90\%C.L.$ |
| Cheng & Tseng | $0.045 \pm 0.011$ | | |
| Żenczykowski | 0.023 | | |
| Datta | 0.01 | | |
| Körner & Krämer | 0.05 | | |